# Rule Based Metadata Extraction Framework from Academic Articles

Jahongir Azimjonov,  aazim0990@gmail.com
Department of Computer and Information
Engineering, Sakarya University, Sakarya, Turkey

Jumabek Alikhanov, jumabek4044@gmail.com
Department of Computer and Information
Engineering, Inha University, Incheon, South Korea

**Abstract-** Metadata of scientific articles such as title, abstract, keywords or index terms, body text, conclusion, reference and others play a decisive role in collecting, managing and storing academic data in scientific databases, academic journals and digital libraries. An accurate extraction of these kinds of data from scientific papers is crucial to organize and retrieve important scientific information for researchers as well as librarians. Research social network systems and academic digital library systems provide academic data extracting, organizing and retrieving services. Mostly these types of services are not free or open source. They also have some performance problems and extracting limitations in the number of PDF (Portable Document Format) files that you can upload to the extraction systems. In this paper, a completely free and open source Java based high performance metadata extraction framework is proposed. This framework's extraction speed is 9-10 times faster than existing metadata extraction systems. It is also flexible in that it allows uploading of unlimited number of PDF files. In this approach, titles of papers are extracted using layout features, font and size characteristics of text. Other metadata fields such as abstracts, body text, keywords, conclusions and references are extracted from PDF files using fixed rule sets. Extracted metadata are stored in both Oracle database and XML (Extensible Markup Language) file. This framework can be used to make scientific collections in digital libraries, online journals, online and offline scientific databases, government research agencies and research centers.

## 1. Introduction

Novel research works, new inventions and modern technologies have been developing rapidly in the last decades. Scientists and researchers publish their recent findings and detailed information about their discoveries in the form of scientific papers in online and offline academic journals or libraries such as ScienceDirect, IEEE XPlore Digital Library, Web of Science, archive.com etc. In addition to this, many institutions, universities, government agencies, research centers etc. are placing their scientific data online. Thus, the amount of academic data on digital network systems is increasing and it is getting more complicated to find specific information day by day. Disarrangement and complexity of data collections make the information search harder and finding desired academic data almost impossible. To overcome these kinds of problems all the scientific papers should be catalogued using their metadata features such as titles, abstracts, index terms, references etc.

It is well known that most of the scientific papers published in digital academic journals or digital libraries are stored as PDF. PDF files contains rich structured and unstructured metadata. To enable scientific articles searchability, metadata fields should be extracted from PDF files and the extracted metadata has to be indexed and saved in databases, XML or JSON (JavaScript Object Notation) files. Extracting metadata from scientific papers is quite easy if the number of files are few and if they consist of structured metadata fields. However, in reality we have to handle large numbers of PDF files whereas these files are not structured.

Manually extracting metadata from large PDF collections is an extremely time-consuming task. According to Chrystal [1], it would take about 60 employee-years to create a metadata collection for 1 million documents. So there is need to automate extraction and collection processes. Developing a software to automate metadata extraction process is a research challenge. However, nothing is impossible. Metadata extraction researches and applications from the past show that metadata can be extracted automatically from PDF files. There are lots of commercial applications which are used widely by digital libraries, online journals, web applications, research social networks and online scientific databases, specifically IEEE XPlore Digital Library, Elsevier, Mendeley, Research Gate, archive.com and "Tubitak Dergi Parki". These systems offer useful academic services to researchers such as publishing, collecting, storing, searching, retrieving scientific data and reference management in research papers. Each of these systems tends to provide better academic services to their customers so that they can have a cutting edge in the competition.

However, these types of academic data management system services are not free. Most of them are not open source, and they have also some performance-speed problems and extracting limitations in the number of PDF files that can be uploaded to the extraction systems. In this approach, a completely free and open source Java based metadata extraction framework with two -three times higher performance, nine-ten times higher speed and unlimited number of PDF files that you may upload in the extraction system is proposed. We use layout and textual features, geometric position, size and font features of text to extract titles of articles and some string and word based fixed rules to extract other metadata fields such as abstracts, body text, keywords, conclusions and references from the PDF files. This framework can be used to make scientific collections in digital libraries, online journals, online and offline scientific databases, government research agencies and centers. Search engines can also utilize this open source framework in their academic network systems. The framework contains a Java package file called "*MetaDataExtractor*" with three distinct classes. The package will be available soon on GitHub (https://github.com). It can be used in metadata extraction projects. The step is easy: just download the package and integrate it into your project. The open source metadata extraction software proposed in this approach was tested with more than six thousand PDF files and compared with other previous and current metadata extraction frameworks. According to the achieved results, it can be concluded that the accuracy and certainty of the results of this framework are better than the expected outcomes.

## 2. Previous Metadata Extraction Approaches.

Metadata extraction from public documents with extensions such as *.pdf, *.docx, *.xls, *.xml, and *.html is theoretically and practically well discovered in the researches [2][3][4][5][6][7][8][9][10][11][12][13][14][15][16][17]. Many of them focus on only specific fields of metadata, which has resulted in them achieving accurate metadata extracting outcomes. Methodologies and techniques that have been used in the recent researches has also resulted in high quality outcomes. Below previous approaches will be discussed in detail.

Previous researches have investigated and applied methods and techniques based on artificial intelligence, learnt statistical measurements, machine-aided indexing or automated indexing, heuristics, machine learning and document specific solutions such as Case-Based Reasoning(CBR) [2], Dublin Core and Bayesian classification [3], rule based [4,7,13] and learning based systems [5,8,9,10], support vector machines[5], supervised and unsupervised learning methods [6], Hidden Markov Model [8], simple heuristic method[11], Hadoop map reduced method[12], hybrid method using GROBID, ParsCit and Mendeley[14], spatial knowledge based system[15], rhetorical sentence classification and extractive text summarization[16] and low level document image processing, and layout analysis [17].

Rajendra et al. [2] presented a content metadata extraction (ME) framework from scientific articles using the case-based reasoning (CBR) method where they emphasized that the most important problem of content extraction from web sources is classification of HTML tag sequences. They argue that accurate classification of HTML tag sequences which contain scientific metadata provide valid content extraction of a scientific paper represented in web documents. CBR is a branch of artificial intelligence that utilizes past specific problem solutions to solve the current problem. The method includes three subtasks; *corpus acquisition, pattern extraction, and content extraction.* The *corpus acquisition* subtask detects a web page of a scientific paper, labels or segments the web page into blocks using HTML tags, then categorizes the labeled HTML tag sequences based on whether they include scientific text or not. The next component, *pattern extraction*, classifies HTML tag sequences as either informative or non-informative patterns. In the last step, *content extraction*, extracts contents of the academic article from the classified patterns of the HTML tag sequences. In CBR approach, they used HTML and its features to extract contents of a scientific paper.

XML based methodologies of ME are also well discovered. For instance, Emma et al. [3] presented a new method to extract keywords and metadata using a "paperBase" automated metadata extraction system which converts a pdf file into a Dublin Core XML file. The system uses Bayesian classifiers to identify metadata in the XML file. Paul et al. [4] shows how XML based methodology can be used to extract metadata from research articles. They use an automated template-based ME architecture which uses *pdftk* framework to split first and last five pages of an article. They convert a pdf article into XML format file using ScanSoft's OmniPage Pro software then they use a rule based system to eliminate metadata from the XML file. In another XML based project proposed by Sukyoung et al. [18], some classification methods have been used to identify text block segmentations and extract metadata from a pdf document by analyzing spatial and layout features of a raw text.

Machine Learning (ML) approaches, specifically Support Vector Machine (SVM) algorithms and rule based systems [5] are other popular methodologies that have been used. Hui et al. [5] used SVM to detect and classify words of each line in a paper. Then they used rule based system method to extract metadata from the clustered words. However, in CERMINE ME framework [6], machine learning supervised and unsupervised learning techniques are used to adopt the system to new previously unseen document layouts and styles. Label segmentation, font size, geometric position and other crucial features of texts of input paper are analyzed and categorized by using ML techniques. After which metadata is extracted from the preprocessed data through rule based and heuristic methods. In another project proposed by Alexandru et al. [7], rule based systems are used to extract tables and some other fields of metadata regardless of formatting style of articles. In the first stage, the fixed rules construct a primary geometric model of tables, images and other fields of metadata. In the second stage, the fixed rules extract metadata from the structured geometric models from the first stage based on special features of metadata fields.

Hidden Markov Models (HMM) and learning based system approaches have also been extensively investigated. Kristie et al. [8] proposed a HMM based framework to eliminate metadata from a research paper. The framework is used to construct certain models from headers of a research paper by labeling each word of a header. The models contain specific states which belong to every single of metadata fields and transitions among the specific states. The constructed models are then used to extract meta information. Aleksandar et al. [9] presented a system based on machine learning approach for extracting meta information. Their approach entails two main tasks and eight subtasks which are responsible for classification and feature extraction of metadata fields of academic articles. In the first task, a PDF file is converted into HTML format file and features of texts are extracted according to characteristics of the first page of a paper. In the second task, the extracted features will be classified as metadata fields using machine learning algorithms such as decision tree, naïve Bayes, K-nearest neighbor and SVM. Xiaonan et al. [10] came up with a system that resembles the project proposed by Aleksandar et al. [9]. However, the differences between two approaches are in file format conversion and classification algorithms that were used. Aleksandar et al. [9] converts a PDF file into HTML document while Xiaonan et al. [10] converts a PDF file into XML file. Furthermore, the approaches presented in [11] and [12] literature have used simple techniques of machine learning to extract titles from scientific articles while Ammar et al. [19] have been utilized font size of a text and rule based heuristic method to detect and pull titles out from research papers.

Moreover, researchers discovered and implemented hybrid methods, frameworks and software tools integrating them to their metadata extraction systems. For example, Tin et al. [13] designed a system integrating several information extraction frameworks and algorithms like a PDFBOX, JAPE and ANNIE tools of GATE framework and rule based system while Ozair et al. [14] proposed a hybrid metadata extraction system integrating GROBID, ParsCit and Mendeley. They emphasized that the metadata extraction process has been divided into tasks. All tasks have been shared between the tools and frameworks. The tools process the raw data and extracts fields of metadata. Giovanni et al. [15] built a visual/spatial knowledge rule based system for metadata extraction from PostScript files. They used GCLIPS framework to detect and eliminate metadata text from a PostScript format file using the fixed 76 rules. However, this approach is not very successful as a text mining based approach proposed by Francesco et al. [16]. This combined approach aimed at extracting metadata from PDF documents by integrating the feature extraction frameworks of the GATE system and the DRI (DR. Inventor) text mining framework. In another research, Simone [17] proposed to develop a ME software called pdf2gsdl which is based on a suitable combination of several techniques that include PDF parsing, low level document image processing and layout analysis. The software contains several modules such as JPedal for PDF parsing and low-level image processing, JDOM for text serialization and obtaining XML file and DBLPCheck for verification of obtained meta information.

3. Metadata and its availability in scientific documents.

Metadata is "data that describes data" in that it contains specific pieces of information in a document or it gives attribute-level description of a document. Any specific document metadata are generated while the documents are created, edited, updated. Metadata is widely used as one of the core features of any software applications on the internet as well as on native applications. Some of the applications of metadata today include: identifying and locating audio and videos on YouTube, searching and listening to music on SoundCloud, writing and sharing a post on Twitter and uploading photos on Instagram. All these applications come with metadata about creation, editing and deleting date, name, topic, and other item features which enables these companies to create customized content and enable development of state of the art recommender systems for their customers. Metadata is also an essential part of any scientific paper because it includes all vital features of an academic article such as title, abstract, authors, affiliation, body text, citations, conclusions, e-mail, keywords or index terms, references, section headings (table of contents) and others which describe virtually any research article. Metadata facilitates accessibility and ease of searching of research articles in digital libraries or online scientific journals. Metadata is vital in providing detailed information about the research which is the reason scientific journals and digital libraries organize their academic services according to the extracted metadata from the pdf files.

There are two main types of pdf files such as structured and unstructured. Structured pdf documents have all filled metadata fields while unstructured pdf files contain only basic metadata features like creation date, file name or page numbers and fonts of a document. Fig.1. shows that metadata can be easily extracted from structured pdf files because all metadata field information in the files are filled properly. Unfortunately, majority of scientific publications lie in the unstructured form due to attention deficiency throughout the creation and editing of the pdf documents which results in unstructured documents' metadata field information being filled poorly complicating their extraction process. It has been proven impossible to extract metadata manually if number of PDF documents involved is large because it is time-consuming. It may take tens of years to pull out metadata manually from thousands of pdf files. However, if the number of files are few, we can effortlessly extract metadata by hand.

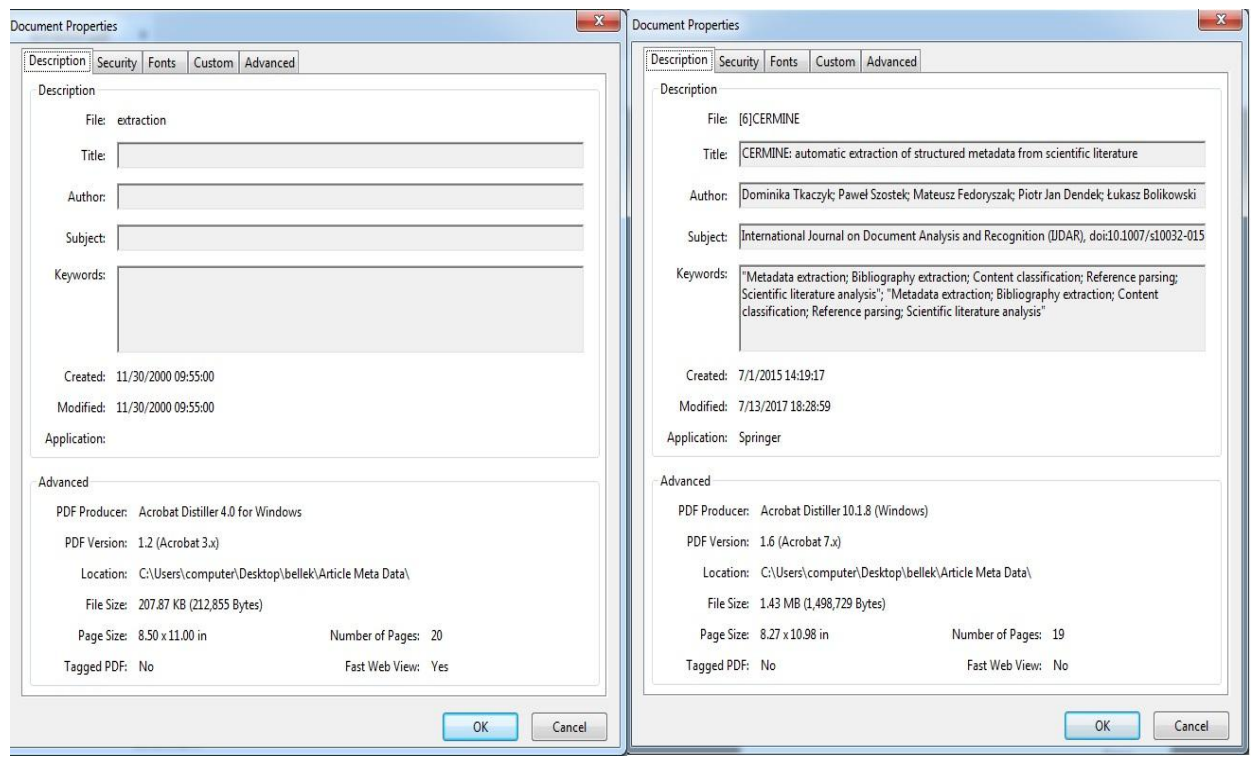

**Fig.1. Example of structured and unstructured form of a pdf document**.

## 4. System architecture of the proposed approach.

The proposed system consists of three main stages: *Collecting and classifying of PDF files*, *Metadata Extracting* and *Indexing and Storing*. Fig.2. shows the whole process in detail. In the first step, research papers are collected from open access scientific journals. For this project, approximately 250 open access computer science journals were identified from which nearly 10,000 PDF documents were downloaded with the assumption that they were research papers using the IDM's (Internet Download Manager) grabber tool. The harvested PDF documents were classified and divided into scientific and unscientific document groups by means of scientific papers' characteristic features explained later in the paper. Then PDF papers are uploaded to the system to apply metadata extraction.

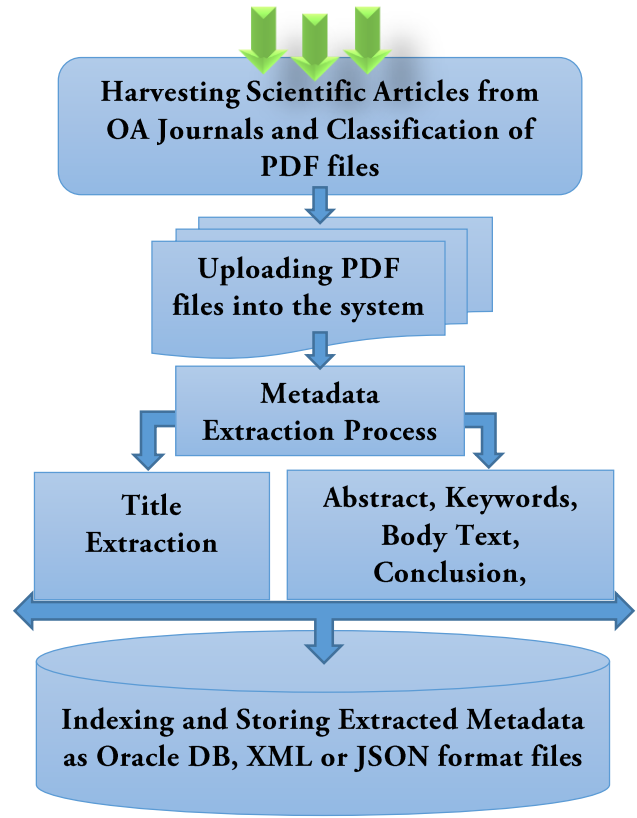

Fig.2. Metadata Extraction System Architecture

The second stage is the core of the system where the first page text with all its font styles, font sizes, layout features as well as the text of last 6 pages are pulled out for the further extraction processes. Text with a bold font and the biggest font size which is between the beginning string and the word "Abstract" or "ABSTRACT" in the first page is predicted to be a title of the input file. Keywords or index terms, abstract, body text, conclusions and references and other metadata fields are extracted word based or layout based fixed field specific rules. If the system is not able to extract any kinds of metadata or if any types of metadata fields in a pdf file are missing, the document will be sent for manual reviewing and the missing metadata is completed manually. In the last stage, harvested metadata is indexed using the name of a PDF file and it can be stored as an Oracle database, XML or JSON file. When any type of metadata is searched, the request is sent to the server which processes the query by searching for appropriate results according to the accepted request through crawling the whole data in the stored files and returns the discovered searching results.

## 5. System implementation

The framework has been developed using Java programming language for the reason that Java provides a wide range of Optical Character Recognition (OCR) libraries and software packages can be integrated to each other effortlessly. Moreover, Java has a very good multiprocessing and multithreading features which increases the computing speed of the framework. Java was chosen for development of the framework due to these functionalities and flexibilities. The framework contains three classes. The first class and its methods do preprocessing by uploading PDF documents to the system and classifying each document if it is scientific article or not. The preprocessed data is uploaded to the extraction module (second class) which is responsible for extracting the desired metadata from PDF files. The third class takes over from the extraction module by indexing each file with its metadata and storing all the extracted data as Oracle Database, XML or JSON file.

### 5.1. Collecting and classifying of PDF files

This is the first stage of the project. In this step PDF files are searched and harvested using spider and robot programs from the determined open source scientific journals and all the discovered PDF documents which are predicted to be scientific papers are classified according to the feature set of research papers. The biggest font size texts, key expressions in the first page such as "Abstract or ABSTRACT" and "Keywords or Index Terms" and key phrases from last pages of a document like "Conclusions or CONCLUSIONS", "REFERENCES or References" are considered as features of scientific papers. If any input document includes these features, the document is categorized as a scientific article and it is forwarded to the metadata extraction process.

### 5.2. Metadata Extracting

This stage is the main and central part of the framework because it consists of metadata extracting class and its methods. The system accepts documents one by one, it pulls out the texts with their layout, font and font size properties from the first page and the last two pages if the total pages of the input file lesser than 7 pages or it takes the four pages if they are greater than seven or 9 pages. The PDFBOX Java software package is used to extract texts with required features and these harvested rich texts are utilized in the further extracting processes.

The *title* method of the second class extracts the *title* of an input document using the text features or in another words three fixed rules: the first rule is layout checking, this rule checks whether the target text which is predicted as the title of the input document lies between the first line and the key phrase "Abstract or ABSTRACT" of the first page text or not; the second one is whether the target text is bold or not; the third rule is whether the font size of the target text is the biggest among the font sizes of first page texts. If all the conditions are met the target text is considered as the title of the incoming document.

The *abstract* method of the class checks whether the predicted text resides between the key phrases "Abstract or ABSTRACT" and "KEYWORDS or Keywords or INDEX TERMS or Index Terms" or not. If the condition is met, the predicted text is taken as *abstract* metadata of the incoming document.

The *keywords* method extracts a text as *keywords* or *index terms* if the text lies down between "KEYWORDS or Keywords or INDEX TERMS or Index Terms" and "I. Intro, 1. Intro or Intro".

The *bodytext* method gets the text as *body* text metadata between key phrases "I. Intro, 1. Intro or Intro" and "Conclusion or CONCLUSION".

The *conclusion* method pulls a text out as *conclusion* metadata if the text lays between key phrases "Conclusion or CONCLUSION" and "Reference or REFERENCES or ACKNOWLADGMENT or Acknowledgement".

The *reference* method takes all text as *reference* metadata of the incoming PDF file starting the keyword "REFERENCE or Reference".

*5.3. Indexing and Storing*

The third stage is responsible for indexing the input PDF files and all the extracted metadata information as Oracle database, XML or JSON format files. The indexed documents and their metadata can be integrated effortlessly to any type of search engines or digital libraries systems.

## 6. Evaluation and Results

To evaluate the framework, 10,000 PDF files were downloaded from the computer science open access journals. 6387 of the 10000 PDF files were real academic articles and the remaining 3613 PDF files were not scientific documents. Firstly, these documents were preprocessed (classified) and the classified pdf files were uploaded to the system for the further extracting processes. Then the extracted metadata results of the current project were compared to outcomes of the previous metadata extractor tools such as GROBID, PDFX, ParsCit, Mendeley Desktop and SciPlore Xtract.

This project has two different outcomes: document classification results (categorizing all input data whether they are scientific paper or not) and metadata extraction results. In the classification process, all 10,000 documents are categorized into two groups namely scientific and unscientific papers. The classifier algorithm recognized 6493 documents as scientific paper and 3507 files as unscientific paper with an accuracy of 97.71% for the classification algorithm. The accuracy has been measured using the formula A= (A1+A2)/2, where A1 is the percentage of truly recognized scientific papers and A2 is the percentage of truly recognized unscientific papers. The pdf documents categorized as scientific papers have been used in the further metadata extraction processes. Firstly, these classified documents were uploaded to the extraction and reliable extraction results for each metadata field were obtained. Secondly, previous metadata extraction tools are tested with the same input data (the same PDF documents) and good results were obtained too which were compared with the results of this research's framework. The comparison results are presented in Table.1. Accuracy for each of the metadata fields is calculated using the three test outcomes and each test has its specific number of input files respectively 100, 1000, 6387. For every single metadata field, accuracy of each test result was calculated and used to assess the value of overall accuracy for each metadata field. As seen in results from Table.1, the overall accuracy for the title is 91.21%, Abstract is 98.13%, Keywords or Index Terms is 92.53%, Body Text is 99.37%, Conclusions is 96.63% and References is 100%.

**Table.1. Metadata extraction results**

| Tools /Metadata | MetaDataExtractor (%) | GROBID (%) | PDFX (%) | ParsCit (%) | Mendeley Desktop (%) | SciPlore Xtract (%) |
|---|---|---|---|---|---|---|
| **Title** | 91.21 | 87.09 | 88.73 | 55.30 | 89.33 | 79.41 |
| **Abstract** | 98.13 | 83.89 | 91.25 | 25.79 | 77.85 | N/A |
| **Keywords** | 92.53 | 90.74 | 57.98 | N/A | 87.89 | N/A |
| **Body Text** | 99.37 | 96.12 | N/A | N/A | N/A | N/A |
| **Conclusions** | 96.63 | N/A | 92.5 | N/A | N/A | N/A |
| **References** | 100 | 95.32 | 97.66 | 75.94 | 96.59 | N/A |

## 7. Conclusion

In this paper a faster, free, flexible Java based open source metadata extraction framework is proposed. The framework contains three classes which does collection and classification of pdf data, extraction of metadata and indexing-storing input pdf files. PDF files from different open access computer science journals were collected and classified as scientific or unscientific. Then all classified scientific pdf files were forwarded to extraction module to pull out the six different metadata fields from each input file. At the end of the process, all pdf documents and their extracted metadata were indexed and stored. The framework spends 3 to 5 seconds to extract 6 different metadata from one pdf file. Overall, the framework extracts, indexes and stores 15 to 20 pdf files per minute. The extraction speed of the framework is 9 or 10 times faster than current metadata extraction software packages. An accuracy for scientific paper classification of the framework is 97.71 % while metadata extraction accuracy results for the title is 91.21%, for the abstract is 98.13%, for the keywords or index terms is 92.53%, for the body text is 99.37%, for the conclusions is 96.63% and for the references is 100%. According to the achieved results, it can be concluded that the accuracy and certainty of the results of this framework are better than the results of the software applications used in the industry. However, this framework extract only 6 metadata fields which is below par considering there are more than 20 fields that can be extracted. As a future work we are planning to add other metadata fields to our framework and we have a plan to develop the framework as a web application.